\documentclass[superscriptaddress,pra,reprint,showpacs]{revtex4-1}

\usepackage{graphicx}
\usepackage{amsmath}
\usepackage{verbatim}
\usepackage{textcomp}
\usepackage{latexsym}
\usepackage{amssymb}
\usepackage{color}

\begin{document}

%\title{Demonstration of Memory-built-in Optical One-way Quantum Computing}
\title{Demonstration of active feedforward one-way quantum computing with photon-matter hyperentanglement}

\author{Xiao-Fan Xu}
\affiliation{Physikalisches Institut, Ruprecht-Karls-Universit\"at Heidelberg, Philosophenweg 12, 69120 Heidelberg, Germany}

\author{Xiao-Hui Bao}
\affiliation{Physikalisches Institut, Ruprecht-Karls-Universit\"at Heidelberg, Philosophenweg 12, 69120 Heidelberg, Germany}
\affiliation{Hefei National Laboratory for Physical Sciences at
Microscale and Department of Modern Physics, University of Science
and Technology of China, Hefei, Anhui 230026, China}

\author{Jian-Wei Pan}
\affiliation{Physikalisches Institut, Ruprecht-Karls-Universit\"at Heidelberg, Philosophenweg 12, 69120 Heidelberg, Germany}
\affiliation{Hefei National Laboratory for Physical Sciences at
Microscale and Department of Modern Physics, University of Science
and Technology of China, Hefei, Anhui 230026, China}

\begin{abstract}
We report an optical one-way quantum computing experiment with stationary quantum memory involved. First we create a hybrid four-qubit cluster state with two qubits propagating as photons and the other two stationary and stored in a laser-cooled atomic-ensemble quantum memory, and characterize it with entanglement witness and quantum state tomography. Then, by making use of this cluster state and fast operations of electro-optic modulators, we realize memory-assisted feedforward operations and demonstrate deterministic single-qubit rotation as an example.
\end{abstract}


\maketitle

%************************* Text ******************************
Quantum computing offers tremendous speedup of certain computing tasks such as factorization of large numbers \cite{SHOR:1994:ID368}, database searching \cite{GROVER:1997:ID166}, simulating quantum systems \cite{Feynman1982, Buluta2009} etc. Due to the ultraweak coupling with the environment, ease of high-precision single-qubit manipulation for all degrees of freedom, and potential for a high experimental repetition rate, optical quantum computing has attracted extensive interest \cite{Kok2007, O'Brien2009}. There have been a number of significant experimental achievements in recent years, both in the circuit \cite{Kok2007} and in the one-way architecture \cite{WALTHER:2005:ID305, CHEN:2007:ID176, Tame2007, PREVEDEL:2007:ID338, VALLONE:2008:ID344, Tokunaga2008, Kaltenbaek2010}, which have proved that optical quantum computing can work in principle. However, due to the probabilistic character of photon sources \cite{Scheel2009, Shields2007} and entangling operations \cite{Kok2007}, efficient optical quantum computing is considered hard to achieve \cite{Browne2005} without making use of quantum memories \cite{Lvovsky2009, Simon2010, Sangouard2011}.

In optical quantum computing, the role of quantum memory is to store photonic qubits such that operations can be timed appropriately \cite{Lvovsky2009, Kok2007}. For instance, in the creation \cite{Browne2005} of large cluster states \cite{BRIEGEL:2001:ID307}, quantum memory is required to store intermediate entangled states while waiting for offline preparation of auxiliary resources, which is essential to making the creation process efficient. Specifically, in order to realize active feedforward operations in a one-way quantum computer \cite{RAUSSENDORF:2001:ID306, BRIEGEL:2009:ID394}, storage of the remaining qubits with quantum memories is necessary so that feedforward operations can be applied according to previous measurement results.

In this Rapid Communication we report an optical one-way quantum computing experiment with stationary quantum memory involved. The major resource is a hyperentangled photon spin-wave cluster state, with the spin-wave stored in a laser-cooled atomic ensemble. The storage capability of the quantum memory helps to realize the necessary active feedforward operations, together with use of fast electro-optic modulators. As an example, a demonstration on deterministic single-qubit rotations is presented in detail. In comparison to previous experiments with active feedforward \cite{PREVEDEL:2007:ID338, VALLONE:2008:ID344}, in which long optical fibers were used, our experiment mainly features flexibility and potential low loss for future large-scale applications.

\begin{figure*}[!tbp]
\centering
\includegraphics[width=0.6\textwidth] {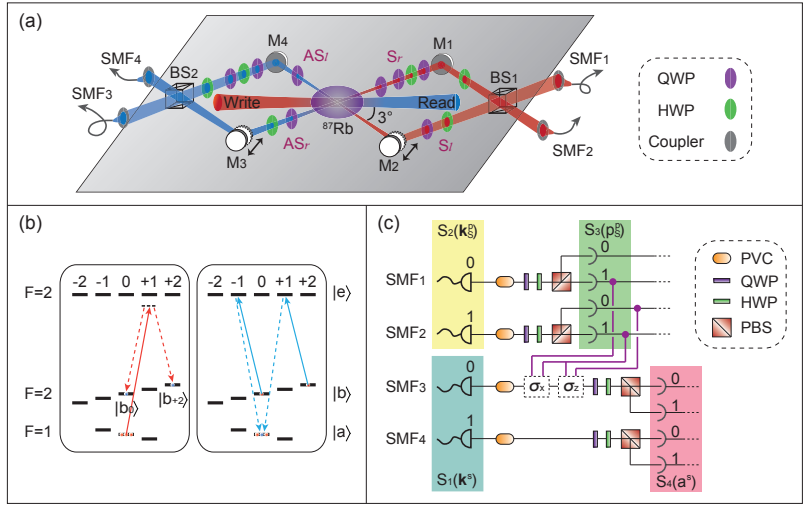}
\caption[]{(color online). Schematic view of the experiment. (a) Experimental setup to create four-qubit hybrid entangled state. During the write process, we select two detection modes (S$_l$ or S$_r$) for the Stokes photon to create photon spin-wave entanglement in the spatial degree. All the photonic modes are guided by single-mode fibers (SMFs) to the extended setup. (b) Level scheme utilized. The write beam is 10 MHz red-detuned from the transition $|a\rangle \leftrightarrow |e\rangle$. Double-pathway Raman scattering (to either $|b_0\rangle$ or $|b_{+2}\rangle$) is utilized to create photon spin-wave entanglement in the polarization degree. (c) Extended setup of (a) to realize active one-way quantum computing. Measurement of the polarization degree of the Stokes photon is carried out by using a combination of a quarter-wave plate (QWP), a half-wave plate (HWP), and a polarizing beam splitter (PBS), while the spatial degree is measured with BS1 and the mirror M2 with piezoelectric ceramics. The spin-wave state is measured through conversion to an anti-Stokes photon and making similar measurements as for the Stokes photon. Pumping vapor cells (PVCs) are used to filter out leakage of the write and read beams. Two electro-optic modulators (EOMs) performing Pauli error corrections are essential for deterministic one-way quantum computation. The overall detection efficiency for the Stokes (anti-Stokes) photon is $\sim0.25$ ($\sim0.20$), including propagation efficiency, coupling efficiency, detector efficiency etc.
}
\label{setup}
\end{figure*}

%\textbf{Preparation of a hybrid four-qubit cluster state.}
A cluster state \cite{BRIEGEL:2001:ID307} serves as the major resource for one-way quantum computing \cite{RAUSSENDORF:2001:ID306}. In our experiment, the cluster is shared between a single photon and an atomic ensemble which serves as the quantum memory. In previous experimental studies, different methods have been used to create photon-memory entanglement, such as using the interference of different spatial modes \cite{CHEN:2007:ID62} or of different Raman pathways \cite{MATSUKEVICH:2005:ID210}. To create a hyperentangled four-qubit state, we make use of both methods simultaneously. The experimental setup is shown in Fig. \ref{setup}. An atomic ensemble of about $10^8$ atoms of \textsuperscript{87}Rb is loaded via a magneto-optical trap and prepared initially in the state $|5^2 S_{1/2}, F=1,m_{F}=0\rangle$. During the write process, through Raman scattering, a spin-wave excitation is created either in the state of $|b_0\rangle$ accompanied by a $\sigma^+$-polarized Stokes photon, or in the state of $|b_{+2}\rangle$ accompanied by a $\sigma^-$-polarized Stokes photon. The Stokes photon is collected either in the spatial mode of $l$ (S$_l$ in Fig. \ref{setup}) with the spin-wave wave vector in the direction of $\downarrow$, or in the spatial mode of $r$(S$_r$ in Fig. \ref{setup}) with the spin-wave wave vector in the direction of $\uparrow$. This twofold correlation enables us to create a hyperentangled photon spin-wave state in the form of
\begin{equation*}
|\psi \rangle = \frac{1}{2} (|\sigma^+\rangle_{\textrm{S}} |b_0\rangle + |\sigma^-\rangle_{\textrm{S}} |b_{+2}\rangle) \otimes (|l\rangle_{\textrm{S}}|\downarrow\rangle + e^{i\theta} |r\rangle_{\textrm{S}} |\uparrow\rangle),
\end{equation*}
where state vectors with a subscript of $\textrm{S}$ correspond to the states of the scattered Stokes photon, state vectors without any subscript correspond to the spin-wave states of the atomic ensemble, and $\theta$ is the propagating phase difference between two spatial modes before they overlap at nonpolarizing beam splitter (BS), and could be compensated by moving a mirror (M2) with piezoelectric ceramics (PIEZO). We rotate the Stokes polarization from $\sigma^+-\sigma^-$ basis to $H-V$ basis with a quarter-wave plate (QWP). In order to create a highly entangled cluster state, we introduce a $\pi$ phase shift between the $V$ and $H$ polarizations in the spatial mode of S$_r$ for the Stokes photon with a combination of two QWPs and a half-wave plate (HWP). Therefore, we get the desired cluster state in the form
\begin{align*}
|C_4 \rangle = \frac{1}{2} ( & |Hl\rangle_{\textrm{S}} |b_0\downarrow\rangle + |Vl\rangle_{\textrm{S}} |b_{+2}\downarrow\rangle \\
 + &|Hr\rangle_{\textrm{S}} |b_0\uparrow\rangle - |Vr\rangle_{\textrm{S}} |b_{+2}\uparrow\rangle) \\
\equiv \frac{1}{2} ( & |0\rangle_{\textrm 1}|0\rangle_{\textrm 2}|0\rangle_{\textrm 3}|0\rangle_{\textrm 4} + |1\rangle_{\textrm 1}|0\rangle_{\textrm 2}|1\rangle_{\textrm 3}|0\rangle_{\textrm 4} \\
 + & |0\rangle_{\textrm 1}|1\rangle_{\textrm 2}|0\rangle_{\textrm 3}|1\rangle_{\textrm 4} - |1\rangle_{\textrm 1}|1\rangle_{\textrm 2}|1\rangle_{\textrm 3}|1\rangle_{\textrm 4}),
\end{align*}
in which we encode logical qubits as $|H,V\rangle_{\textrm S} \leftrightarrow |0,1\rangle_{\textrm 1}$, $|l,r\rangle_{\textrm S} \leftrightarrow |0,1\rangle_{\textrm 2}$, $|b_0,b_{+2}\rangle \leftrightarrow |0,1\rangle_{\textrm 3}$, and $|\downarrow,\uparrow\rangle \leftrightarrow |0,1\rangle_{\textrm 4}$.

%\textbf{Characterization of the hybrid cluster state.}
The cluster state is first evaluated by an optimal entanglement witness via the stabilizer operators \cite{KIESEL:2005:ID361}. The witness is of the form
\begin{align*}
\mathcal{W}= \frac{1}{2} & [4I^{\otimes 4} -(\sigma_x I \sigma_x \sigma_z + \sigma_x \sigma_z \sigma_x I + I \sigma_z I \sigma_z \\
& + I \sigma_x \sigma_z \sigma_x + \sigma_z \sigma_x I \sigma_x + \sigma_z I \sigma_z I)],
\end{align*}
where $I$ is a two-dimensional identity matrix, while $\sigma_z = |0\rangle\langle0|-|1\rangle\langle1|$, $\sigma_x = |0\rangle\langle1|+|1\rangle\langle0|$ are Pauli matrices. A negative value of the witness indicates existence of quadripartite entanglement. The minimum value of -1 refers to an ideal cluster state of $|C_4\rangle$. We measure the Stokes polarization with a combination of a QWP, a HWP, and a polarizing beam splitter (PBS). For the spatial degree measurement of the Stokes photon, we use a BS and PIEZO-mounted mirror (e.g., M2) for the basis $(|0\rangle \pm e^{i \alpha} |1\rangle)/\sqrt{2}$, or remove the BS for the basis $|0\rangle,|1\rangle$. In order to measure the spin-wave states, we apply a read pulse to efficiently convert the spin-wave to an anti-Stokes photon. Thus, the spin-wave states are converted to the corresponding photonic states as $|b_0,b_{+2}\rangle \leftrightarrow |H,V\rangle_{\textrm AS}$ and $|\downarrow , \uparrow \rangle \leftrightarrow |l,r\rangle_{\textrm AS}$, which could be measured in the same way as the Stokes photon. In the case of 2.27 $\mu$s spin-wave storage, we obtain $\langle \mathcal{W} \rangle_{exp} = -0.60 \pm 0.01$ which clearly demonstrates the genuine quadripartite entanglement, and yields a lower-bound for the fidelity $F \ge \frac{1}{2} - \frac{1}{2} \langle \mathcal{W} \rangle_{exp} = 0.800 \pm 0.007$.

\begin{figure}[!tbp]
\centering
\includegraphics[width=0.35\textwidth] {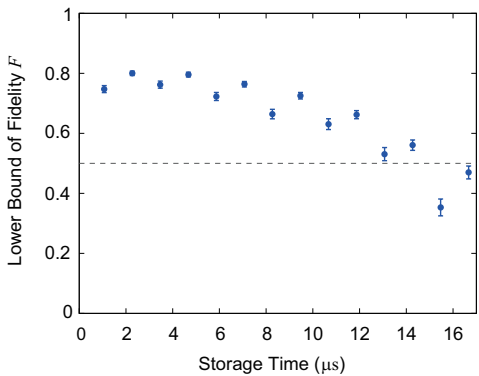}
\caption[]{(color online). Measured fidelity of the cluster state as a function of storage time. The fidelity values shown in the figure are the lower-bound values obtained from the witness measurements, which indicate a lifetime of the cluster state of about 14.27 $\mu$s. Error bars represent statistical errors.
}
\label{lifetime}
\end{figure}

To demonstrate the storage capability of this hybrid cluster state, we measure the entanglement witness for various storage durations and present the results in Fig. \ref{lifetime}. The point where the fidelity is larger than 0.5 \cite{SACKETT:2000:ID386} indicates that the lifetime is longer than 14.27 $\mu$s. The dephasing of the spin-wave induced by atomic random motion is the principal mechanism that limits the lifetime, and it can be weakened by decreasing the detection angle to increase the wavelength of the spin-wave \cite{ZHAO:2009:ID56}. In the present experiment, we use the angle of $3^{\circ}$ between the direction of the write pulse and the direction of the Stokes field. The fidelity oscillation in Fig. \ref{lifetime} is due to imperfect optical pumping which gives rise to slight collapse and revival for the read process \cite{Matsukevich2006}.

In order to evaluate the cluster state further, we reconstruct the density matrix of the cluster by quantum state tomography via maximum-likelihood technique \cite{JAMES:2001:ID237}, and calculate the fidelity directly from the reconstructed state to be $0.817 \pm 0.004$ which agrees with the witness result very well. In order to investigate the reason for the imperfect cluster state preparation, we calculate the reduced density matrices for each degree. In comparison with the desired entangled state of $1/\sqrt{2}(|H\rangle_{\textrm{S}} |b_0\rangle \pm |V\rangle_{\textrm{S}} |b_{+2}\rangle$ for the polarization degree, and $1/\sqrt{2}(|l\rangle_{\textrm{S}} |\downarrow\rangle \pm |r\rangle_{\textrm{S}} |\uparrow\rangle)$ for the spatial degree, we get an average fidelity of 88.5(5)\% and 95.5(4)\%, respectively. The relatively low fidelity for the polarization degree is mainly caused by the imbalance of the strength of the corresponding transitions used in the write process. Nonideal entanglement preparation in the spatial degree is mainly caused by higher-order excitations and the imperfection of the BSs.

\begin{figure}[!tbp]
\centering
\includegraphics[width=0.45\textwidth] {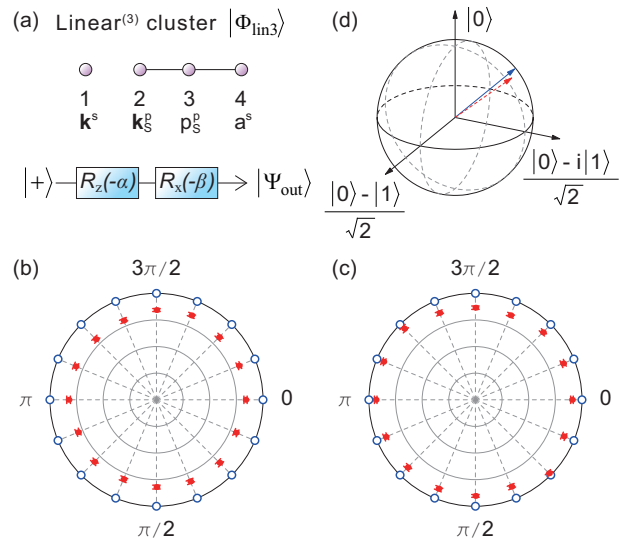}
\caption[]{(color online). Demonstration of arbitrary single-qubit rotations. (a) The one-way quantum circuit configuration of single-qubit rotations. (b), (c) Experimental results for the fidelities, which are represented by distances to the circle center. In (b), $\alpha = \pi/2$ and $\beta$ is set from $0$ to $2 \pi$ in steps of $\pi/8$. In (c), $\alpha$ is set from $0$ to $2 \pi$ in steps of $\pi/8$ and $\beta = 0$. The blue open dots show the ideal states with fidelity of 1, and each of the red dots with error bars shows the experimental result for the corresponding ideal state located in the same dashed line. The four concentric circles mark fidelities of 0.25, 0.5, 0.75, and 1 sequentially from the center. (d) Bloch sphere representation of the tomography results for the rotation operation $R_x(- \pi / 4) R_z(- \pi /4)$ on logical qubit $|+\rangle$. The blue arrow describes the ideal state, while the red dashed arrow describes the measured state obtained from the reconstructed density matrix. The fidelity is calculated as $0.93 \pm 0.02$.}
\label{rotation}
\end{figure}

%\textbf{One-way quantum computing.}
This hybrid cluster state enables us to demonstrate active one-way quantum computing with built-in quantum memory. As an example, we make a proof-of-principle demonstration of deterministic single-qubit rotations by using quantum-memory-assisted active feedforward. In one-way quantum computing \cite{RAUSSENDORF:2001:ID306}, implementation of a certain quantum circuit requires a cluster state of a certain pattern. In order to perform arbitrary single-qubit rotations, a three-qubit linear cluster state $|\Phi_{\textrm{lin3}}\rangle$ as shown in Fig. \ref{rotation} is required. In our experiment, we get such a cluster by following three steps: (a) rearranging the order of the four qubits as $(1,2,3,4) = (\textrm{\textbf{k}}^{\textrm{s}},\textrm{\textbf{k}}^{\textrm {p}}_{\textrm{S}},{\textrm{p}}^{\textrm{p}}_\textrm{S},{\textrm{a}}^\textrm{s})$
with the notation of ${\textrm{p}}^{\textrm{p}}_\textrm{S}$ for the polarization degree of the Stokes photon, $\textrm{\textbf{k}}^{\textrm {p}}_{\textrm{S}}$ for the spatial degree of the Stokes photon, ${\textrm{a}}^\textrm{s}$ for the polarization degree of the spin-wave and $\textrm{\textbf{k}}^{\textrm{s}}$ for spatial degree of the spin-wave;
(b) implementing a unitary operator $H \otimes I \otimes I \otimes H$ on our experimental state, with $H = (\sigma_x + \sigma_z)/\sqrt{2}$ in which $\sigma_x$, $\sigma_y$, $\sigma_z$ are Pauli matrices, $I$ being the identity matrix;
(c) removing the first qubit $\textrm{\textbf{k}}^{\textrm{s}}$ through postselective measurement in the computational basis $\{|0\rangle_1, |1\rangle_1\}$ \cite{notea}. In our experiment, the Hadamard transformation $H$ on qubit 1 ($\textrm{\textbf{k}}^{\textrm{s}}$) is performed by BS2, while the Hadamard transformation $H$ on qubit 4 (${\textrm{a}}^\textrm{s}$) is performed by setting the last HWP in the two retrieved photon paths at $67.5^{\circ}$.

Arbitrary single-qubit rotation can be realized through measuring qubit 2 and 3 in the bases $B_2(\alpha)$ and $B_3(\beta)$ consecutively, where $B_j(\alpha) = \{|\alpha_+\rangle_j,|\alpha_-\rangle_j\}$ with $|\alpha_{\pm}\rangle_j = (|0\rangle_j \pm e^{i \alpha}|1\rangle_j)/\sqrt{2}$. The effective rotation applied onto the encoded qubit can be expressed as:
\begin{equation}
\label{eqrotation}
|\Psi_{\textrm{out}}\rangle = \sigma_x^{s_3} \sigma_z^{s_2} R_x((-1)^{s_2+1}\beta) R_z(-\alpha) |\Psi_{\textrm{in}}\rangle,
\end{equation}
where $R_{x,z}(\alpha)=\textrm{exp}(-i \alpha \sigma_{x,z} /2)$, $|\Psi_{\textrm{in}}\rangle = |+\rangle = 1/\sqrt{2}(|0\rangle + |1\rangle)$ \cite{notea}, and $s_i$ takes the value of 0 or 1 corresponding to the outcome of the measurement on qubit $i$. Eq. (\ref{eqrotation}) indicates that the measurement basis $B_3(\pm\beta)$ of qubit 3 depends on the previous measurement outcome $s_2$ of qubit 2 (type-I error), and the random measurement outcomes will induce random Pauli errors for the logic qubits (type-II error). In order to make the single-qubit rotation deterministic, both types of the error have to be corrected which necessitates an active feedforward technique. One of the most advisable methods is to take advantage of flexible quantum storage together with fast switchable optical elements such as electro-optic modulators (EOMs). Quantum storage of successive qubits compensates the time delay of the preceding measurements as well as the response time of the EOMs which are utilized to change measurement bases and apply Pauli error corrections actively. For the single-qubit rotation in Eq. (\ref{eqrotation}), correction of type-I error requests feedforward between different degrees of freedom, namely, from $\textrm{\textbf{k}}^{\textrm {p}}_{\textrm{S}}$ to ${\textrm{p}}^{\textrm{p}}_\textrm{S}$, where EOMs are not necessary. Specifically, as shown in Fig. \ref{setup}(a), measurement $B_2(\alpha)$ is carried out by BS1 with the aid of M2, where different measurement outcomes $s_2$ correspond to different output modes of BS1. Thus, feedforward from qubit 2 to 3 can be simply realized by making polarization measurements in different bases $B_3(\pm\beta)$ for different output modes of BS1 ($s_2=0$ or $s_2=1$).

In order to correct type-II errors in Eq. (\ref{eqrotation}), the measurement outcomes of $s_2$ and $s_3$ have to be actively fed-forward onto qubit 4. This feedforward process includes single-photon detection to get $s_2$ and $s_3$, conversion of qubit 4 from a spin-wave to a single photon, and application of Pauli error corrections onto the photonic state with two EOMs based on the outcomes $s_2$ and $s_3$. The total time delay of this process is about 1.69 $\mu$s, in which the response time of the EOM system contributes 1.56 $\mu$s, optical propagation contributes 20 ns, and electrical signal processing and propagation contributes 110 ns. To compensate this time delay, we make use of the storage capability of the atomic ensemble to store qubit 4. The lifetime (14.27 $\mu$s) of the spin-wave state is much longer than the total time delay, thus supporting much longer time delay or more steps of feedforward operations. In this single-qubit gate experiment, qubit 4 is stored in the atomic ensemble for 2.27 $\mu$s before it is read out (with an efficiency of $\sim$ 0.29) and subjected to Pauli error corrections.

In Fig. \ref{rotation}, we show the results of single-qubit rotation $R_x(-\beta)$ and $R_z(-\alpha)$ from $0$ to $2 \pi$ in step of $\pi/8$. The average fidelity is $0.82 \pm 0.02$ for $R_x(-\beta)$ rotations shown in Fig. \ref{rotation}(b), and $0.91 \pm 0.03$ for $R_z(-\alpha)$ rotations shown in Fig. \ref{rotation}(c). We attribute this difference mainly to the characteristics of our cluster state. $R_z(-\alpha)$ rotations correspond to measurements of qubit 2 ($\textrm{\textbf{k}}^{\textrm {p}}_{\textrm{S}}$) in various bases, referring to the spatial degree with better preparation quality; while $R_z(-\beta)$ rotations correspond to the measurements of qubit 3 (${\textrm{p}}^{\textrm{p}}_\textrm{S}$) in various bases, referring to the polarization degree with lower preparation quality. Moreover, in order to show the ability for arbitrary rotations, we choose a set of specific values of $\alpha = \pi/4$ and $\beta = \pi/4$. The desired and measured target states are plotted in the Bloch sphere shown in Fig. \ref{rotation}(d), with a calculated fidelity of $0.93 \pm 0.02$. Excluding the errors induced during the cluster state preparation, we attribute the leftover error sources to the imperfection of the EOMs and other optical elements used.

%\section{Discussion}
To conclude, we have created a four-qubit cluster state which is shared between a single-photon and a spin-wave state of an atomic ensemble. This hybrid cluster state enables us to realize optical one-way quantum computing with built-in quantum memory. As an example, we have demonstrated deterministic single-qubit rotations by making use of fast EOMs and the storage capability of the atomic ensemble. In comparison with previous purely photonic demonstrations of one-way quantum computing, our experiment provides additional experimental capabilities. The atomic ensemble not only acts as an integral quantum memory
to store part of the cluster state, but also enables some extent of tunability of the frequency \cite{Yuan2007}, pulse duration \cite{Bao2012}, and readout time point for the converted single photon, which is advantageous for further connection with other physical systems in the context of quantum networks \cite{Kimble2008}. In the current experiment, the maximum allowable number of steps of active feedforward is $\sim 7$, which is mainly limited by the slow rise time of the EOM driver and the short spin-wave coherence time. If faster EOM drivers \cite{PREVEDEL:2007:ID338} (65 $\mathrm{ns}$ rise time) are employed, and an optical lattice is utilized to confine the motion of individual atoms, which leads to a spin-wave coherence time of 100 $\mathrm{ms}$ \cite{Radnaev2010}, realization of $10^6$ steps of feedforward is foreseeable in the near future. Considering the similarity between a spin-wave excitation and a single photon, it is possible to encode even more information on one atomic ensemble, e.g., by selecting more spatial modes \cite{Rossi2009}, using the angular momentum degree \cite{Inoue2009}, including more ground states, etc. While this method of using more degrees of freedom to create larger cluster states is not scalable in general, a scheme \cite{Barrett2010} recently proposed by Barrett \textsl{et al.} claims scalable generation of cluster states involving more photons and more memory units.

We thank C.-Y. Lu, Y.-A. Chen, B. Zhao and Z.-S. Yuan for helpful discussions. This work was supported by the European Commission through the ERC Grant, the STREP project HIP, the CAS, the NNSFC, and the National Fundamental Research Program (Grant No. 2011CB921300) of China.

\bibliography{myref}

\begin{thebibliography}{37}%
\makeatletter
\providecommand \@ifxundefined [1]{%
 \@ifx{#1\undefined}
}%
\providecommand \@ifnum [1]{%
 \ifnum #1\expandafter \@firstoftwo
 \else \expandafter \@secondoftwo
 \fi
}%
\providecommand \@ifx [1]{%
 \ifx #1\expandafter \@firstoftwo
 \else \expandafter \@secondoftwo
 \fi
}%
\providecommand \natexlab [1]{#1}%
\providecommand \enquote  [1]{``#1''}%
\providecommand \bibnamefont  [1]{#1}%
\providecommand \bibfnamefont [1]{#1}%
\providecommand \citenamefont [1]{#1}%
\providecommand \href@noop [0]{\@secondoftwo}%
\providecommand \href [0]{\begingroup \@sanitize@url \@href}%
\providecommand \@href[1]{\@@startlink{#1}\@@href}%
\providecommand \@@href[1]{\endgroup#1\@@endlink}%
\providecommand \@sanitize@url [0]{\catcode `\\12\catcode `\$12\catcode
  `\&12\catcode `\#12\catcode `\^12\catcode `\_12\catcode `\%12\relax}%
\providecommand \@@startlink[1]{}%
\providecommand \@@endlink[0]{}%
\providecommand \url  [0]{\begingroup\@sanitize@url \@url }%
\providecommand \@url [1]{\endgroup\@href {#1}{\urlprefix }}%
\providecommand \urlprefix  [0]{URL }%
\providecommand \Eprint [0]{\href }%
\providecommand \doibase [0]{http://dx.doi.org/}%
\providecommand \selectlanguage [0]{\@gobble}%
\providecommand \bibinfo  [0]{\@secondoftwo}%
\providecommand \bibfield  [0]{\@secondoftwo}%
\providecommand \translation [1]{[#1]}%
\providecommand \BibitemOpen [0]{}%
\providecommand \bibitemStop [0]{}%
\providecommand \bibitemNoStop [0]{.\EOS\space}%
\providecommand \EOS [0]{\spacefactor3000\relax}%
\providecommand \BibitemShut  [1]{\csname bibitem#1\endcsname}%
\let\auto@bib@innerbib\@empty
%</preamble>
\bibitem [{\citenamefont {Shor}(1994)}]{SHOR:1994:ID368}%
  \BibitemOpen
  \bibfield  {author} {\bibinfo {author} {\bibfnamefont {P.~W.}\ \bibnamefont
  {Shor}},\ }in\ \href@noop {} {\emph {\bibinfo {booktitle} {Proc. 35th Annu.
  Symp. Foundations of Computer Science}}},\ \bibinfo {editor} {edited by\
  \bibinfo {editor} {\bibfnamefont {S.}~\bibnamefont {Goldwasser}}}\ (\bibinfo
  {publisher} {IEEE Computer Society Press},\ \bibinfo {address} {Los
  Alamitos},\ \bibinfo {year} {1994})\ pp.\ \bibinfo {pages}
  {124--134}\BibitemShut {NoStop}%
\bibitem [{\citenamefont {Grover}(1997)}]{GROVER:1997:ID166}%
  \BibitemOpen
  \bibfield  {author} {\bibinfo {author} {\bibfnamefont {L.~K.}\ \bibnamefont
  {Grover}},\ }\href@noop {} {\bibfield  {journal} {\bibinfo  {journal} {Phys.
  Rev. Lett.}\ }\textbf {\bibinfo {volume} {79}},\ \bibinfo {pages} {325}
  (\bibinfo {year} {1997})}\BibitemShut {NoStop}%
\bibitem [{\citenamefont {Feynman}(1982)}]{Feynman1982}%
  \BibitemOpen
  \bibfield  {author} {\bibinfo {author} {\bibfnamefont {R.}~\bibnamefont
  {Feynman}},\ }\href@noop {} {\bibfield  {journal} {\bibinfo  {journal} {Int.
  J. Theor. Phys.}\ }\textbf {\bibinfo {volume} {21}},\ \bibinfo {pages} {467}
  (\bibinfo {year} {1982})}\BibitemShut {NoStop}%
\bibitem [{\citenamefont {Buluta}\ and\ \citenamefont
  {Nori}(2009)}]{Buluta2009}%
  \BibitemOpen
  \bibfield  {author} {\bibinfo {author} {\bibfnamefont {I.}~\bibnamefont
  {Buluta}}\ and\ \bibinfo {author} {\bibfnamefont {F.}~\bibnamefont {Nori}},\
  }\href@noop {} {\bibfield  {journal} {\bibinfo  {journal} {Science}\ }\textbf
  {\bibinfo {volume} {326}},\ \bibinfo {pages} {108} (\bibinfo {year}
  {2009})}\BibitemShut {NoStop}%
\bibitem [{\citenamefont {Kok}\ \emph {et~al.}(2007)\citenamefont {Kok},
  \citenamefont {Munro}, \citenamefont {Nemoto}, \citenamefont {Ralph},
  \citenamefont {Dowling},\ and\ \citenamefont {Milburn}}]{Kok2007}%
  \BibitemOpen
  \bibfield  {author} {\bibinfo {author} {\bibfnamefont {P.}~\bibnamefont
  {Kok}}, \bibinfo {author} {\bibfnamefont {W.~J.}\ \bibnamefont {Munro}},
  \bibinfo {author} {\bibfnamefont {K.}~\bibnamefont {Nemoto}}, \bibinfo
  {author} {\bibfnamefont {T.~C.}\ \bibnamefont {Ralph}}, \bibinfo {author}
  {\bibfnamefont {J.~P.}\ \bibnamefont {Dowling}}, \ and\ \bibinfo {author}
  {\bibfnamefont {G.~J.}\ \bibnamefont {Milburn}},\ }\href@noop {} {\bibfield
  {journal} {\bibinfo  {journal} {Rev. Mod. Phys.}\ }\textbf {\bibinfo {volume}
  {79}},\ \bibinfo {pages} {135} (\bibinfo {year} {2007})}\BibitemShut
  {NoStop}%
\bibitem [{\citenamefont {O'Brien}\ \emph {et~al.}(2009)\citenamefont
  {O'Brien}, \citenamefont {Furusawa},\ and\ \citenamefont
  {Vuckovic}}]{O'Brien2009}%
  \BibitemOpen
  \bibfield  {author} {\bibinfo {author} {\bibfnamefont {J.~L.}\ \bibnamefont
  {O'Brien}}, \bibinfo {author} {\bibfnamefont {A.}~\bibnamefont {Furusawa}}, \
  and\ \bibinfo {author} {\bibfnamefont {J.}~\bibnamefont {Vuckovic}},\
  }\href@noop {} {\bibfield  {journal} {\bibinfo  {journal} {Nat. Photonics}\
  }\textbf {\bibinfo {volume} {3}},\ \bibinfo {pages} {687} (\bibinfo {year}
  {2009})}\BibitemShut {NoStop}%
\bibitem [{\citenamefont {Walther}\ \emph {et~al.}(2005)\citenamefont
  {Walther}, \citenamefont {Resch}, \citenamefont {Rudolph}, \citenamefont
  {Schenck}, \citenamefont {Weinfurter}, \citenamefont {Vedral}, \citenamefont
  {Aspelmeyer},\ and\ \citenamefont {Zeilinger}}]{WALTHER:2005:ID305}%
  \BibitemOpen
  \bibfield  {author} {\bibinfo {author} {\bibfnamefont {P.}~\bibnamefont
  {Walther}}, \bibinfo {author} {\bibfnamefont {K.~J.}\ \bibnamefont {Resch}},
  \bibinfo {author} {\bibfnamefont {T.}~\bibnamefont {Rudolph}}, \bibinfo
  {author} {\bibfnamefont {E.}~\bibnamefont {Schenck}}, \bibinfo {author}
  {\bibfnamefont {H.}~\bibnamefont {Weinfurter}}, \bibinfo {author}
  {\bibfnamefont {V.}~\bibnamefont {Vedral}}, \bibinfo {author} {\bibfnamefont
  {M.}~\bibnamefont {Aspelmeyer}}, \ and\ \bibinfo {author} {\bibfnamefont
  {A.}~\bibnamefont {Zeilinger}},\ }\href@noop {} {\bibfield  {journal}
  {\bibinfo  {journal} {Nature}\ }\textbf {\bibinfo {volume} {434}},\ \bibinfo
  {pages} {169} (\bibinfo {year} {2005})}\BibitemShut {NoStop}%
\bibitem [{\citenamefont {Chen}\ \emph
  {et~al.}(2007{\natexlab{a}})\citenamefont {Chen}, \citenamefont {Li},
  \citenamefont {Zhang}, \citenamefont {Chen}, \citenamefont {Goebel},
  \citenamefont {Chen}, \citenamefont {Mair},\ and\ \citenamefont
  {Pan}}]{CHEN:2007:ID176}%
  \BibitemOpen
  \bibfield  {author} {\bibinfo {author} {\bibfnamefont {K.}~\bibnamefont
  {Chen}}, \bibinfo {author} {\bibfnamefont {C.-M.}\ \bibnamefont {Li}},
  \bibinfo {author} {\bibfnamefont {Q.}~\bibnamefont {Zhang}}, \bibinfo
  {author} {\bibfnamefont {Y.-A.}\ \bibnamefont {Chen}}, \bibinfo {author}
  {\bibfnamefont {A.}~\bibnamefont {Goebel}}, \bibinfo {author} {\bibfnamefont
  {S.}~\bibnamefont {Chen}}, \bibinfo {author} {\bibfnamefont {A.}~\bibnamefont
  {Mair}}, \ and\ \bibinfo {author} {\bibfnamefont {J.-W.}\ \bibnamefont
  {Pan}},\ }\href@noop {} {\bibfield  {journal} {\bibinfo  {journal} {Phys.
  Rev. Lett.}\ }\textbf {\bibinfo {volume} {99}},\ \bibinfo {pages} {120503}
  (\bibinfo {year} {2007}{\natexlab{a}})}\BibitemShut {NoStop}%
\bibitem [{\citenamefont {Tame}\ \emph {et~al.}(2007)\citenamefont {Tame},
  \citenamefont {Prevedel}, \citenamefont {Paternostro}, \citenamefont {Bohi},
  \citenamefont {Kim},\ and\ \citenamefont {Zeilinger}}]{Tame2007}%
  \BibitemOpen
  \bibfield  {author} {\bibinfo {author} {\bibfnamefont {M.~S.}\ \bibnamefont
  {Tame}}, \bibinfo {author} {\bibfnamefont {R.}~\bibnamefont {Prevedel}},
  \bibinfo {author} {\bibfnamefont {M.}~\bibnamefont {Paternostro}}, \bibinfo
  {author} {\bibfnamefont {P.}~\bibnamefont {Bohi}}, \bibinfo {author}
  {\bibfnamefont {M.~S.}\ \bibnamefont {Kim}}, \ and\ \bibinfo {author}
  {\bibfnamefont {A.}~\bibnamefont {Zeilinger}},\ }\href@noop {} {\bibfield
  {journal} {\bibinfo  {journal} {Phys. Rev. Lett.}\ }\textbf {\bibinfo
  {volume} {98}},\ \bibinfo {pages} {140501} (\bibinfo {year}
  {2007})}\BibitemShut {NoStop}%
\bibitem [{\citenamefont {Prevedel}\ \emph {et~al.}(2007)\citenamefont
  {Prevedel}, \citenamefont {Walther}, \citenamefont {Tiefenbacher},
  \citenamefont {Bohi}, \citenamefont {Kaltenbaek}, \citenamefont {Jennewein},\
  and\ \citenamefont {Zeilinger}}]{PREVEDEL:2007:ID338}%
  \BibitemOpen
  \bibfield  {author} {\bibinfo {author} {\bibfnamefont {R.}~\bibnamefont
  {Prevedel}}, \bibinfo {author} {\bibfnamefont {P.}~\bibnamefont {Walther}},
  \bibinfo {author} {\bibfnamefont {F.}~\bibnamefont {Tiefenbacher}}, \bibinfo
  {author} {\bibfnamefont {P.}~\bibnamefont {Bohi}}, \bibinfo {author}
  {\bibfnamefont {R.}~\bibnamefont {Kaltenbaek}}, \bibinfo {author}
  {\bibfnamefont {T.}~\bibnamefont {Jennewein}}, \ and\ \bibinfo {author}
  {\bibfnamefont {A.}~\bibnamefont {Zeilinger}},\ }\href@noop {} {\bibfield
  {journal} {\bibinfo  {journal} {Nature}\ }\textbf {\bibinfo {volume} {445}},\
  \bibinfo {pages} {65} (\bibinfo {year} {2007})}\BibitemShut {NoStop}%
\bibitem [{\citenamefont {Vallone}\ \emph {et~al.}(2008)\citenamefont
  {Vallone}, \citenamefont {Pomarico}, \citenamefont {De~Martini},\ and\
  \citenamefont {Mataloni}}]{VALLONE:2008:ID344}%
  \BibitemOpen
  \bibfield  {author} {\bibinfo {author} {\bibfnamefont {G.}~\bibnamefont
  {Vallone}}, \bibinfo {author} {\bibfnamefont {E.}~\bibnamefont {Pomarico}},
  \bibinfo {author} {\bibfnamefont {F.}~\bibnamefont {De~Martini}}, \ and\
  \bibinfo {author} {\bibfnamefont {P.}~\bibnamefont {Mataloni}},\ }\href@noop
  {} {\bibfield  {journal} {\bibinfo  {journal} {Phys. Rev. Lett.}\ }\textbf
  {\bibinfo {volume} {100}},\ \bibinfo {pages} {160502} (\bibinfo {year}
  {2008})}\BibitemShut {NoStop}%
\bibitem [{\citenamefont {Tokunaga}\ \emph {et~al.}(2008)\citenamefont
  {Tokunaga}, \citenamefont {Kuwashiro}, \citenamefont {Yamamoto},
  \citenamefont {Koashi},\ and\ \citenamefont {Imoto}}]{Tokunaga2008}%
  \BibitemOpen
  \bibfield  {author} {\bibinfo {author} {\bibfnamefont {Y.}~\bibnamefont
  {Tokunaga}}, \bibinfo {author} {\bibfnamefont {S.}~\bibnamefont {Kuwashiro}},
  \bibinfo {author} {\bibfnamefont {T.}~\bibnamefont {Yamamoto}}, \bibinfo
  {author} {\bibfnamefont {M.}~\bibnamefont {Koashi}}, \ and\ \bibinfo {author}
  {\bibfnamefont {N.}~\bibnamefont {Imoto}},\ }\href@noop {} {\bibfield
  {journal} {\bibinfo  {journal} {Phys. Rev. Lett.}\ }\textbf {\bibinfo
  {volume} {100}},\ \bibinfo {pages} {210501} (\bibinfo {year}
  {2008})}\BibitemShut {NoStop}%
\bibitem [{\citenamefont {Kaltenbaek}\ \emph {et~al.}(2010)\citenamefont
  {Kaltenbaek}, \citenamefont {Lavoie}, \citenamefont {Zeng}, \citenamefont
  {Bartlett},\ and\ \citenamefont {Resch}}]{Kaltenbaek2010}%
  \BibitemOpen
  \bibfield  {author} {\bibinfo {author} {\bibfnamefont {R.}~\bibnamefont
  {Kaltenbaek}}, \bibinfo {author} {\bibfnamefont {J.}~\bibnamefont {Lavoie}},
  \bibinfo {author} {\bibfnamefont {B.}~\bibnamefont {Zeng}}, \bibinfo {author}
  {\bibfnamefont {S.~D.}\ \bibnamefont {Bartlett}}, \ and\ \bibinfo {author}
  {\bibfnamefont {K.~J.}\ \bibnamefont {Resch}},\ }\href@noop {} {\bibfield
  {journal} {\bibinfo  {journal} {Nat. Phys.}\ }\textbf {\bibinfo {volume}
  {6}},\ \bibinfo {pages} {850} (\bibinfo {year} {2010})}\BibitemShut {NoStop}%
\bibitem [{\citenamefont {Scheel}(2009)}]{Scheel2009}%
  \BibitemOpen
  \bibfield  {author} {\bibinfo {author} {\bibfnamefont {S.}~\bibnamefont
  {Scheel}},\ }\href@noop {} {\bibfield  {journal} {\bibinfo  {journal} {J.
  Mod. Opt.}\ }\textbf {\bibinfo {volume} {56}},\ \bibinfo {pages} {141}
  (\bibinfo {year} {2009})}\BibitemShut {NoStop}%
\bibitem [{\citenamefont {Shields}(2007)}]{Shields2007}%
  \BibitemOpen
  \bibfield  {author} {\bibinfo {author} {\bibfnamefont {A.~J.}\ \bibnamefont
  {Shields}},\ }\href@noop {} {\bibfield  {journal} {\bibinfo  {journal} {Nat.
  Photonics}\ }\textbf {\bibinfo {volume} {1}},\ \bibinfo {pages} {215}
  (\bibinfo {year} {2007})}\BibitemShut {NoStop}%
\bibitem [{\citenamefont {Browne}\ and\ \citenamefont
  {Rudolph}(2005)}]{Browne2005}%
  \BibitemOpen
  \bibfield  {author} {\bibinfo {author} {\bibfnamefont {D.~E.}\ \bibnamefont
  {Browne}}\ and\ \bibinfo {author} {\bibfnamefont {T.}~\bibnamefont
  {Rudolph}},\ }\href@noop {} {\bibfield  {journal} {\bibinfo  {journal} {Phys.
  Rev. Lett.}\ }\textbf {\bibinfo {volume} {95}},\ \bibinfo {pages} {010501}
  (\bibinfo {year} {2005})}\BibitemShut {NoStop}%
\bibitem [{\citenamefont {Lvovsky}\ \emph {et~al.}(2009)\citenamefont
  {Lvovsky}, \citenamefont {Sanders},\ and\ \citenamefont
  {Tittel}}]{Lvovsky2009}%
  \BibitemOpen
  \bibfield  {author} {\bibinfo {author} {\bibfnamefont {A.~I.}\ \bibnamefont
  {Lvovsky}}, \bibinfo {author} {\bibfnamefont {B.~C.}\ \bibnamefont
  {Sanders}}, \ and\ \bibinfo {author} {\bibfnamefont {W.}~\bibnamefont
  {Tittel}},\ }\href@noop {} {\bibfield  {journal} {\bibinfo  {journal} {Nat.
  Photonics}\ }\textbf {\bibinfo {volume} {3}},\ \bibinfo {pages} {706}
  (\bibinfo {year} {2009})}\BibitemShut {NoStop}%
\bibitem [{\citenamefont {Simon}\ \emph {et~al.}(2010)\citenamefont {Simon},
  \citenamefont {Afzelius}, \citenamefont {Appel}, \citenamefont {Boyer de~la
  Giroday}, \citenamefont {Dewhurst}, \citenamefont {Gisin}, \citenamefont
  {Hu}, \citenamefont {Jelezko}, \citenamefont {Kr\"oll}, \citenamefont
  {M\"uller}, \citenamefont {Nunn}, \citenamefont {Polzik}, \citenamefont
  {Rarity}, \citenamefont {De~Riedmatten}, \citenamefont {Rosenfeld},
  \citenamefont {Shields}, \citenamefont {Sk\"old}, \citenamefont {Stevenson},
  \citenamefont {Thew}, \citenamefont {Walmsley}, \citenamefont {Weber},
  \citenamefont {Weinfurter}, \citenamefont {Wrachtrup},\ and\ \citenamefont
  {Young}}]{Simon2010}%
  \BibitemOpen
  \bibfield  {author} {\bibinfo {author} {\bibfnamefont {C.}~\bibnamefont
  {Simon}}, \bibinfo {author} {\bibfnamefont {M.}~\bibnamefont {Afzelius}},
  \bibinfo {author} {\bibfnamefont {J.}~\bibnamefont {Appel}}, \bibinfo
  {author} {\bibfnamefont {A.}~\bibnamefont {Boyer de~la Giroday}}, \bibinfo
  {author} {\bibfnamefont {S.~J.}\ \bibnamefont {Dewhurst}}, \bibinfo {author}
  {\bibfnamefont {N.}~\bibnamefont {Gisin}}, \bibinfo {author} {\bibfnamefont
  {C.~Y.}\ \bibnamefont {Hu}}, \bibinfo {author} {\bibfnamefont
  {F.}~\bibnamefont {Jelezko}}, \bibinfo {author} {\bibfnamefont
  {S.}~\bibnamefont {Kr\"oll}}, \bibinfo {author} {\bibfnamefont {J.~H.}\
  \bibnamefont {M\"uller}}, \bibinfo {author} {\bibfnamefont {J.}~\bibnamefont
  {Nunn}}, \bibinfo {author} {\bibfnamefont {E.~S.}\ \bibnamefont {Polzik}},
  \bibinfo {author} {\bibfnamefont {J.~G.}\ \bibnamefont {Rarity}}, \bibinfo
  {author} {\bibfnamefont {H.}~\bibnamefont {De~Riedmatten}}, \bibinfo {author}
  {\bibfnamefont {W.}~\bibnamefont {Rosenfeld}}, \bibinfo {author}
  {\bibfnamefont {A.~J.}\ \bibnamefont {Shields}}, \bibinfo {author}
  {\bibfnamefont {N.}~\bibnamefont {Sk\"old}}, \bibinfo {author} {\bibfnamefont
  {R.~M.}\ \bibnamefont {Stevenson}}, \bibinfo {author} {\bibfnamefont
  {R.}~\bibnamefont {Thew}}, \bibinfo {author} {\bibfnamefont {I.~A.}\
  \bibnamefont {Walmsley}}, \bibinfo {author} {\bibfnamefont {M.~C.}\
  \bibnamefont {Weber}}, \bibinfo {author} {\bibfnamefont {H.}~\bibnamefont
  {Weinfurter}}, \bibinfo {author} {\bibfnamefont {J.}~\bibnamefont
  {Wrachtrup}}, \ and\ \bibinfo {author} {\bibfnamefont {R.~J.}\ \bibnamefont
  {Young}},\ }\href@noop {} {\bibfield  {journal} {\bibinfo  {journal} {Eur.
  Phys. J. D}\ }\textbf {\bibinfo {volume} {58}},\ \bibinfo {pages} {1}
  (\bibinfo {year} {2010})}\BibitemShut {NoStop}%
\bibitem [{\citenamefont {Sangouard}\ \emph {et~al.}(2011)\citenamefont
  {Sangouard}, \citenamefont {Simon}, \citenamefont {de~Riedmatten},\ and\
  \citenamefont {Gisin}}]{Sangouard2011}%
  \BibitemOpen
  \bibfield  {author} {\bibinfo {author} {\bibfnamefont {N.}~\bibnamefont
  {Sangouard}}, \bibinfo {author} {\bibfnamefont {C.}~\bibnamefont {Simon}},
  \bibinfo {author} {\bibfnamefont {H.}~\bibnamefont {de~Riedmatten}}, \ and\
  \bibinfo {author} {\bibfnamefont {N.}~\bibnamefont {Gisin}},\ }\href@noop {}
  {\bibfield  {journal} {\bibinfo  {journal} {Rev. Mod. Phys.}\ }\textbf
  {\bibinfo {volume} {83}},\ \bibinfo {pages} {33} (\bibinfo {year}
  {2011})}\BibitemShut {NoStop}%
\bibitem [{\citenamefont {Briegel}\ and\ \citenamefont
  {Raussendorf}(2001)}]{BRIEGEL:2001:ID307}%
  \BibitemOpen
  \bibfield  {author} {\bibinfo {author} {\bibfnamefont {H.~J.}\ \bibnamefont
  {Briegel}}\ and\ \bibinfo {author} {\bibfnamefont {R.}~\bibnamefont
  {Raussendorf}},\ }\href@noop {} {\bibfield  {journal} {\bibinfo  {journal}
  {Phys. Rev. Lett.}\ }\textbf {\bibinfo {volume} {86}},\ \bibinfo {pages}
  {910} (\bibinfo {year} {2001})}\BibitemShut {NoStop}%
\bibitem [{\citenamefont {Raussendorf}\ and\ \citenamefont
  {Briegel}(2001)}]{RAUSSENDORF:2001:ID306}%
  \BibitemOpen
  \bibfield  {author} {\bibinfo {author} {\bibfnamefont {R.}~\bibnamefont
  {Raussendorf}}\ and\ \bibinfo {author} {\bibfnamefont {H.~J.}\ \bibnamefont
  {Briegel}},\ }\href@noop {} {\bibfield  {journal} {\bibinfo  {journal} {Phys.
  Rev. Lett.}\ }\textbf {\bibinfo {volume} {86}},\ \bibinfo {pages} {5188}
  (\bibinfo {year} {2001})}\BibitemShut {NoStop}%
\bibitem [{\citenamefont {Briegel}\ \emph {et~al.}(2009)\citenamefont
  {Briegel}, \citenamefont {Browne}, \citenamefont {D\"ur}, \citenamefont
  {Raussendorf},\ and\ \citenamefont {Van~den Nest}}]{BRIEGEL:2009:ID394}%
  \BibitemOpen
  \bibfield  {author} {\bibinfo {author} {\bibfnamefont {H.~J.}\ \bibnamefont
  {Briegel}}, \bibinfo {author} {\bibfnamefont {D.~E.}\ \bibnamefont {Browne}},
  \bibinfo {author} {\bibfnamefont {W.}~\bibnamefont {D\"ur}}, \bibinfo
  {author} {\bibfnamefont {R.}~\bibnamefont {Raussendorf}}, \ and\ \bibinfo
  {author} {\bibfnamefont {M.}~\bibnamefont {Van~den Nest}},\ }\href@noop {}
  {\bibfield  {journal} {\bibinfo  {journal} {Nat. Phys.}\ }\textbf {\bibinfo
  {volume} {5}},\ \bibinfo {pages} {19} (\bibinfo {year} {2009})}\BibitemShut
  {NoStop}%
\bibitem [{\citenamefont {Chen}\ \emph
  {et~al.}(2007{\natexlab{b}})\citenamefont {Chen}, \citenamefont {Chen},
  \citenamefont {Zhao}, \citenamefont {Yuan}, \citenamefont {Schmiedmayer},\
  and\ \citenamefont {Pan}}]{CHEN:2007:ID62}%
  \BibitemOpen
  \bibfield  {author} {\bibinfo {author} {\bibfnamefont {S.}~\bibnamefont
  {Chen}}, \bibinfo {author} {\bibfnamefont {Y.-A.}\ \bibnamefont {Chen}},
  \bibinfo {author} {\bibfnamefont {B.}~\bibnamefont {Zhao}}, \bibinfo {author}
  {\bibfnamefont {Z.-S.}\ \bibnamefont {Yuan}}, \bibinfo {author}
  {\bibfnamefont {J.}~\bibnamefont {Schmiedmayer}}, \ and\ \bibinfo {author}
  {\bibfnamefont {J.-W.}\ \bibnamefont {Pan}},\ }\href@noop {} {\bibfield
  {journal} {\bibinfo  {journal} {Phys. Rev. Lett.}\ }\textbf {\bibinfo
  {volume} {99}},\ \bibinfo {pages} {180505} (\bibinfo {year}
  {2007}{\natexlab{b}})}\BibitemShut {NoStop}%
\bibitem [{\citenamefont {Matsukevich}\ \emph {et~al.}(2005)\citenamefont
  {Matsukevich}, \citenamefont {Chaneli\`ere}, \citenamefont {Bhattacharya},
  \citenamefont {Lan}, \citenamefont {Jenkins}, \citenamefont {Kennedy},\ and\
  \citenamefont {Kuzmich}}]{MATSUKEVICH:2005:ID210}%
  \BibitemOpen
  \bibfield  {author} {\bibinfo {author} {\bibfnamefont {D.~N.}\ \bibnamefont
  {Matsukevich}}, \bibinfo {author} {\bibfnamefont {T.}~\bibnamefont
  {Chaneli\`ere}}, \bibinfo {author} {\bibfnamefont {M.}~\bibnamefont
  {Bhattacharya}}, \bibinfo {author} {\bibfnamefont {S.-Y.}\ \bibnamefont
  {Lan}}, \bibinfo {author} {\bibfnamefont {S.~D.}\ \bibnamefont {Jenkins}},
  \bibinfo {author} {\bibfnamefont {T.~A.~B.}\ \bibnamefont {Kennedy}}, \ and\
  \bibinfo {author} {\bibfnamefont {A.}~\bibnamefont {Kuzmich}},\ }\href@noop
  {} {\bibfield  {journal} {\bibinfo  {journal} {Phys. Rev. Lett.}\ }\textbf
  {\bibinfo {volume} {95}},\ \bibinfo {pages} {040405} (\bibinfo {year}
  {2005})}\BibitemShut {NoStop}%
\bibitem [{\citenamefont {Kiesel}\ \emph {et~al.}(2005)\citenamefont {Kiesel},
  \citenamefont {Schmid}, \citenamefont {Weber}, \citenamefont {T\'{o}th},
  \citenamefont {G\"{u}hne}, \citenamefont {Ursin},\ and\ \citenamefont
  {Weinfurter}}]{KIESEL:2005:ID361}%
  \BibitemOpen
  \bibfield  {author} {\bibinfo {author} {\bibfnamefont {N.}~\bibnamefont
  {Kiesel}}, \bibinfo {author} {\bibfnamefont {C.}~\bibnamefont {Schmid}},
  \bibinfo {author} {\bibfnamefont {U.}~\bibnamefont {Weber}}, \bibinfo
  {author} {\bibfnamefont {G.}~\bibnamefont {T\'{o}th}}, \bibinfo {author}
  {\bibfnamefont {O.}~\bibnamefont {G\"{u}hne}}, \bibinfo {author}
  {\bibfnamefont {R.}~\bibnamefont {Ursin}}, \ and\ \bibinfo {author}
  {\bibfnamefont {H.}~\bibnamefont {Weinfurter}},\ }\href@noop {} {\bibfield
  {journal} {\bibinfo  {journal} {Phys. Rev. Lett.}\ }\textbf {\bibinfo
  {volume} {95}},\ \bibinfo {pages} {210502} (\bibinfo {year}
  {2005})}\BibitemShut {NoStop}%
\bibitem [{\citenamefont {Sackett}\ \emph {et~al.}(2000)\citenamefont
  {Sackett}, \citenamefont {Kielpinski}, \citenamefont {King}, \citenamefont
  {Langer}, \citenamefont {Meyer}, \citenamefont {Myatt}, \citenamefont {Rowe},
  \citenamefont {Turchette}, \citenamefont {Itano}, \citenamefont {Wineland},\
  and\ \citenamefont {Monroe}}]{SACKETT:2000:ID386}%
  \BibitemOpen
  \bibfield  {author} {\bibinfo {author} {\bibfnamefont {C.~A.}\ \bibnamefont
  {Sackett}}, \bibinfo {author} {\bibfnamefont {D.}~\bibnamefont {Kielpinski}},
  \bibinfo {author} {\bibfnamefont {B.~E.}\ \bibnamefont {King}}, \bibinfo
  {author} {\bibfnamefont {C.}~\bibnamefont {Langer}}, \bibinfo {author}
  {\bibfnamefont {V.}~\bibnamefont {Meyer}}, \bibinfo {author} {\bibfnamefont
  {C.~J.}\ \bibnamefont {Myatt}}, \bibinfo {author} {\bibfnamefont
  {M.}~\bibnamefont {Rowe}}, \bibinfo {author} {\bibfnamefont {Q.~A.}\
  \bibnamefont {Turchette}}, \bibinfo {author} {\bibfnamefont {W.~M.}\
  \bibnamefont {Itano}}, \bibinfo {author} {\bibfnamefont {D.~J.}\ \bibnamefont
  {Wineland}}, \ and\ \bibinfo {author} {\bibfnamefont {C.}~\bibnamefont
  {Monroe}},\ }\href@noop {} {\bibfield  {journal} {\bibinfo  {journal}
  {Nature}\ }\textbf {\bibinfo {volume} {404}},\ \bibinfo {pages} {256}
  (\bibinfo {year} {2000})}\BibitemShut {NoStop}%
\bibitem [{\citenamefont {Zhao}\ \emph {et~al.}(2009)\citenamefont {Zhao},
  \citenamefont {Chen}, \citenamefont {Bao}, \citenamefont {Strassel},
  \citenamefont {Chuu}, \citenamefont {Jin}, \citenamefont {Schmiedmayer},
  \citenamefont {Yuan}, \citenamefont {Chen},\ and\ \citenamefont
  {Pan}}]{ZHAO:2009:ID56}%
  \BibitemOpen
  \bibfield  {author} {\bibinfo {author} {\bibfnamefont {B.}~\bibnamefont
  {Zhao}}, \bibinfo {author} {\bibfnamefont {Y.-A.}\ \bibnamefont {Chen}},
  \bibinfo {author} {\bibfnamefont {X.-H.}\ \bibnamefont {Bao}}, \bibinfo
  {author} {\bibfnamefont {T.}~\bibnamefont {Strassel}}, \bibinfo {author}
  {\bibfnamefont {C.-S.}\ \bibnamefont {Chuu}}, \bibinfo {author}
  {\bibfnamefont {X.-M.}\ \bibnamefont {Jin}}, \bibinfo {author} {\bibfnamefont
  {J.}~\bibnamefont {Schmiedmayer}}, \bibinfo {author} {\bibfnamefont {Z.-S.}\
  \bibnamefont {Yuan}}, \bibinfo {author} {\bibfnamefont {S.}~\bibnamefont
  {Chen}}, \ and\ \bibinfo {author} {\bibfnamefont {J.-W.}\ \bibnamefont
  {Pan}},\ }\href@noop {} {\bibfield  {journal} {\bibinfo  {journal} {Nat.
  Phys.}\ }\textbf {\bibinfo {volume} {5}},\ \bibinfo {pages} {95} (\bibinfo
  {year} {2009})}\BibitemShut {NoStop}%
\bibitem [{\citenamefont {Matsukevich}\ \emph {et~al.}(2006)\citenamefont
  {Matsukevich}, \citenamefont {Chaneli\`{e}re}, \citenamefont {Jenkins},
  \citenamefont {Lan}, \citenamefont {Kennedy},\ and\ \citenamefont
  {Kuzmich}}]{Matsukevich2006}%
  \BibitemOpen
  \bibfield  {author} {\bibinfo {author} {\bibfnamefont {D.~N.}\ \bibnamefont
  {Matsukevich}}, \bibinfo {author} {\bibfnamefont {T.}~\bibnamefont
  {Chaneli\`{e}re}}, \bibinfo {author} {\bibfnamefont {S.~D.}\ \bibnamefont
  {Jenkins}}, \bibinfo {author} {\bibfnamefont {S.-Y.}\ \bibnamefont {Lan}},
  \bibinfo {author} {\bibfnamefont {T.~A.~B.}\ \bibnamefont {Kennedy}}, \ and\
  \bibinfo {author} {\bibfnamefont {A.}~\bibnamefont {Kuzmich}},\ }\href@noop
  {} {\bibfield  {journal} {\bibinfo  {journal} {Phys. Rev. Lett.}\ }\textbf
  {\bibinfo {volume} {96}},\ \bibinfo {pages} {033601} (\bibinfo {year}
  {2006})}\BibitemShut {NoStop}%
\bibitem [{\citenamefont {James}\ \emph {et~al.}(2001)\citenamefont {James},
  \citenamefont {Kwiat}, \citenamefont {Munro},\ and\ \citenamefont
  {White}}]{JAMES:2001:ID237}%
  \BibitemOpen
  \bibfield  {author} {\bibinfo {author} {\bibfnamefont {D.~F.~V.}\
  \bibnamefont {James}}, \bibinfo {author} {\bibfnamefont {P.~G.}\ \bibnamefont
  {Kwiat}}, \bibinfo {author} {\bibfnamefont {W.~J.}\ \bibnamefont {Munro}}, \
  and\ \bibinfo {author} {\bibfnamefont {A.~G.}\ \bibnamefont {White}},\
  }\href@noop {} {\bibfield  {journal} {\bibinfo  {journal} {Phys. Rev. A}\
  }\textbf {\bibinfo {volume} {64}},\ \bibinfo {pages} {052312} (\bibinfo
  {year} {2001})}\BibitemShut {NoStop}%
\bibitem [{not()}]{notea}%
  \BibitemOpen
  \href@noop {} {}\bibinfo {note} {For the measurement of the physical qubit
  $\textrm{\textbf{k}}^{\textrm{s}}$, we postselect the outcome of $s_1=0$.
  This postselection introduces a $50\%$ reduction of the state preparation
  efficiency for $|\Psi_{\textrm{in}}\rangle$, while it does not affect the
  deterministic character of the single-qubit gate.}\BibitemShut {Stop}%
\bibitem [{\citenamefont {Yuan}\ \emph {et~al.}(2007)\citenamefont {Yuan},
  \citenamefont {Chen}, \citenamefont {Chen}, \citenamefont {Zhao},
  \citenamefont {Koch}, \citenamefont {Strassel}, \citenamefont {Zhao},
  \citenamefont {Zhu}, \citenamefont {Schmiedmayer},\ and\ \citenamefont
  {Pan}}]{Yuan2007}%
  \BibitemOpen
  \bibfield  {author} {\bibinfo {author} {\bibfnamefont {Z.-S.}\ \bibnamefont
  {Yuan}}, \bibinfo {author} {\bibfnamefont {Y.-A.}\ \bibnamefont {Chen}},
  \bibinfo {author} {\bibfnamefont {S.}~\bibnamefont {Chen}}, \bibinfo {author}
  {\bibfnamefont {B.}~\bibnamefont {Zhao}}, \bibinfo {author} {\bibfnamefont
  {M.}~\bibnamefont {Koch}}, \bibinfo {author} {\bibfnamefont {T.}~\bibnamefont
  {Strassel}}, \bibinfo {author} {\bibfnamefont {Y.}~\bibnamefont {Zhao}},
  \bibinfo {author} {\bibfnamefont {G.-J.}\ \bibnamefont {Zhu}}, \bibinfo
  {author} {\bibfnamefont {J.}~\bibnamefont {Schmiedmayer}}, \ and\ \bibinfo
  {author} {\bibfnamefont {J.-W.}\ \bibnamefont {Pan}},\ }\href@noop {}
  {\bibfield  {journal} {\bibinfo  {journal} {Phys. Rev. Lett.}\ }\textbf
  {\bibinfo {volume} {98}},\ \bibinfo {pages} {180503} (\bibinfo {year}
  {2007})}\BibitemShut {NoStop}%
\bibitem [{\citenamefont {Bao}\ \emph {et~al.}(2012)\citenamefont {Bao},
  \citenamefont {Reingruber}, \citenamefont {Dietrich}, \citenamefont {Rui},
  \citenamefont {Duck}, \citenamefont {Strassel}, \citenamefont {Li},
  \citenamefont {Liu}, \citenamefont {Zhao},\ and\ \citenamefont
  {Pan}}]{Bao2012}%
  \BibitemOpen
  \bibfield  {author} {\bibinfo {author} {\bibfnamefont {X.-H.}\ \bibnamefont
  {Bao}}, \bibinfo {author} {\bibfnamefont {A.}~\bibnamefont {Reingruber}},
  \bibinfo {author} {\bibfnamefont {P.}~\bibnamefont {Dietrich}}, \bibinfo
  {author} {\bibfnamefont {J.}~\bibnamefont {Rui}}, \bibinfo {author}
  {\bibfnamefont {A.}~\bibnamefont {Duck}}, \bibinfo {author} {\bibfnamefont
  {T.}~\bibnamefont {Strassel}}, \bibinfo {author} {\bibfnamefont
  {L.}~\bibnamefont {Li}}, \bibinfo {author} {\bibfnamefont {N.~L.}\
  \bibnamefont {Liu}}, \bibinfo {author} {\bibfnamefont {B.}~\bibnamefont
  {Zhao}}, \ and\ \bibinfo {author} {\bibfnamefont {J.-W.}\ \bibnamefont
  {Pan}},\ }\href@noop {} {\bibfield  {journal} {\bibinfo  {journal} {Nat.
  Phys.}\ }\textbf {\bibinfo {volume} {8}},\ \bibinfo {pages} {517} (\bibinfo
  {year} {2012})}\BibitemShut {NoStop}%
\bibitem [{\citenamefont {Kimble}(2008)}]{Kimble2008}%
  \BibitemOpen
  \bibfield  {author} {\bibinfo {author} {\bibfnamefont {H.~J.}\ \bibnamefont
  {Kimble}},\ }\href@noop {} {\bibfield  {journal} {\bibinfo  {journal}
  {Nature}\ }\textbf {\bibinfo {volume} {453}},\ \bibinfo {pages} {1023}
  (\bibinfo {year} {2008})}\BibitemShut {NoStop}%
\bibitem [{\citenamefont {Radnaev}\ \emph {et~al.}(2010)\citenamefont
  {Radnaev}, \citenamefont {Dudin}, \citenamefont {Zhao}, \citenamefont {Jen},
  \citenamefont {Jenkins}, \citenamefont {Kuzmich},\ and\ \citenamefont
  {Kennedy}}]{Radnaev2010}%
  \BibitemOpen
  \bibfield  {author} {\bibinfo {author} {\bibfnamefont {A.~G.}\ \bibnamefont
  {Radnaev}}, \bibinfo {author} {\bibfnamefont {Y.~O.}\ \bibnamefont {Dudin}},
  \bibinfo {author} {\bibfnamefont {R.}~\bibnamefont {Zhao}}, \bibinfo {author}
  {\bibfnamefont {H.~H.}\ \bibnamefont {Jen}}, \bibinfo {author} {\bibfnamefont
  {S.~D.}\ \bibnamefont {Jenkins}}, \bibinfo {author} {\bibfnamefont
  {A.}~\bibnamefont {Kuzmich}}, \ and\ \bibinfo {author} {\bibfnamefont
  {T.~A.~B.}\ \bibnamefont {Kennedy}},\ }\href@noop {} {\bibfield  {journal}
  {\bibinfo  {journal} {Nat. Phys.}\ }\textbf {\bibinfo {volume} {6}},\
  \bibinfo {pages} {894} (\bibinfo {year} {2010})}\BibitemShut {NoStop}%
\bibitem [{\citenamefont {Rossi}\ \emph {et~al.}(2009)\citenamefont {Rossi},
  \citenamefont {Vallone}, \citenamefont {Chiuri}, \citenamefont {De~Martini},\
  and\ \citenamefont {Mataloni}}]{Rossi2009}%
  \BibitemOpen
  \bibfield  {author} {\bibinfo {author} {\bibfnamefont {A.}~\bibnamefont
  {Rossi}}, \bibinfo {author} {\bibfnamefont {G.}~\bibnamefont {Vallone}},
  \bibinfo {author} {\bibfnamefont {A.}~\bibnamefont {Chiuri}}, \bibinfo
  {author} {\bibfnamefont {F.}~\bibnamefont {De~Martini}}, \ and\ \bibinfo
  {author} {\bibfnamefont {P.}~\bibnamefont {Mataloni}},\ }\href@noop {}
  {\bibfield  {journal} {\bibinfo  {journal} {Phys. Rev. Lett.}\ }\textbf
  {\bibinfo {volume} {102}},\ \bibinfo {pages} {153902} (\bibinfo {year}
  {2009})}\BibitemShut {NoStop}%
\bibitem [{\citenamefont {Inoue}\ \emph {et~al.}(2009)\citenamefont {Inoue},
  \citenamefont {Yonehara}, \citenamefont {Miyamoto}, \citenamefont {Koashi},\
  and\ \citenamefont {Kozuma}}]{Inoue2009}%
  \BibitemOpen
  \bibfield  {author} {\bibinfo {author} {\bibfnamefont {R.}~\bibnamefont
  {Inoue}}, \bibinfo {author} {\bibfnamefont {T.}~\bibnamefont {Yonehara}},
  \bibinfo {author} {\bibfnamefont {Y.}~\bibnamefont {Miyamoto}}, \bibinfo
  {author} {\bibfnamefont {M.}~\bibnamefont {Koashi}}, \ and\ \bibinfo {author}
  {\bibfnamefont {M.}~\bibnamefont {Kozuma}},\ }\href@noop {} {\bibfield
  {journal} {\bibinfo  {journal} {Phys. Rev. Lett.}\ }\textbf {\bibinfo
  {volume} {103}},\ \bibinfo {pages} {110503} (\bibinfo {year}
  {2009})}\BibitemShut {NoStop}%
\bibitem [{\citenamefont {Barrett}\ \emph {et~al.}(2010)\citenamefont
  {Barrett}, \citenamefont {Rohde},\ and\ \citenamefont {Stace}}]{Barrett2010}%
  \BibitemOpen
  \bibfield  {author} {\bibinfo {author} {\bibfnamefont {S.~D.}\ \bibnamefont
  {Barrett}}, \bibinfo {author} {\bibfnamefont {P.~P.}\ \bibnamefont {Rohde}},
  \ and\ \bibinfo {author} {\bibfnamefont {T.~M.}\ \bibnamefont {Stace}},\
  }\href@noop {} {\bibfield  {journal} {\bibinfo  {journal} {New J. Phys.}\
  }\textbf {\bibinfo {volume} {12}},\ \bibinfo {pages} {093032} (\bibinfo
  {year} {2010})}\BibitemShut {NoStop}%
\end{thebibliography}%
\end{document}